# Time-reversal symmetry breaking driven topological phase transition in EuB$_6$

Shun-Ye Gao,[1,2,#] Sheng Xu,[3,4,#] Hang Li,[5,6,#] Chang-Jiang Yi,[1,#] Si-Min Nie,[7] Zhi-Cheng Rao,[1,2] Huan Wang,[3,4] Quan-Xin Hu,[1,2] Xue-Zhi Chen,[2,8] Wen-Hui Fan,[1,2] Jie-Rui Huang,[1,2] Yao-Bo Huang,[9] Nini Pryds,[5] Ming Shi,[6] Zhi-Jun Wang,[1,2] You-Guo Shi,[1,2,10,*] Tian-Long Xia,[3,4,*] Tian Qian,[1,10,*] Hong Ding[1,2,10,11]

[1] *Beijing National Laboratory for Condensed Matter Physics and Institute of Physics, Chinese Academy of Sciences, Beijing, 100190, China*

[2] *University of Chinese Academy of Sciences, Beijing, 100049, China*

[3] *Department of physics, Renmin University of China, Beijing, 100872, China*

[4] *Beijing Key Laboratory of Opto-electronic Functional Materials & Micro-nano Devices, Renmin University of China, Beijing, 100872, China*

[5] *Department of Energy Conversion and Storage, Technical University of Denmark, Fysikvej Building 310, DK-2800 Kgs. Lyngby, Denmark*

[6] *Swiss Light Source, Paul Scherrer Institut, CH-5232 Villigen, Switzerland*

[7] *Department of Materials Science and Engineering, Stanford University, Stanford, California, 94305, USA*

[8] *Shanghai Institute of Applied Physics, Chinese Academy of Sciences, Shanghai, 201800, China*

[9] *Shanghai Synchrotron Radiation Facility, Shanghai Advanced Research Institute, Chinese Academy of Sciences, Shanghai, 201204, China*

[10] *Songshan Lake Materials Laboratory, Dongguan, Guangdong, 523808, China*

[11] *CAS Center for Excellence in Topological Quantum Computation, University of Chinese Academy of Sciences, Beijing, 100049, China*

[#] These authors contributed equally to this work.

[*] Corresponding authors: tqian@iphy.ac.cn, tlxia@ruc.edu.cn, ygshi@iphy.ac.cn

**Abstract**

The interplay between time-reversal symmetry (TRS) and band topology plays a crucial role in topological states of quantum matter. In time-reversal-invariant (TRI) systems, the inversion of spin-degenerate bands with opposite parity leads to nontrivial topological states, such as topological insulators and Dirac semimetals. When the TRS is broken, the exchange field induces spin splitting of the bands. The inversion of a pair of spin-splitting subbands can generate more exotic topological states, such as quantum anomalous Hall insulators and magnetic Weyl semimetals. So far, such topological phase transitions driven by the TRS breaking have not been visualized. In this work, using angle-resolved photoemission spectroscopy, we have demonstrated that the TRS breaking induces a band inversion of a pair of spin-splitting subbands at the TRI points of Brillouin zone in $EuB_6$, when a long-range ferromagnetic order is developed. The dramatic changes in the electronic structure result in a topological phase transition from a TRI ordinary insulator state to a TRS-broken topological semimetal (TSM) state. Remarkably, the magnetic TSM state has an ideal electronic structure, in which the band crossings are located at the Fermi level without any interference from other bands. Our findings not only reveal the topological phase transition driven by the TRS breaking, but also provide an excellent platform to explore novel physical behavior in the magnetic topological states of quantum matter.

# I. INTRODUCTION

Since the discovery of the integer quantum Hall effect, time-reversal symmetry (TRS) has been a key issue in the study of topological states of quantum matter [1-3]. The thinking and research on TRS directly led to the discovery of time-reversal-invariant (TRI) topological insulators with the quantum spin Hall effect [4-9]. In magnetic materials, the TRS breaking induced by magnetic order can produce more exotic topological states. The most famous examples are the realization of quantum anomalous Hall (QAH) effect [10-19] and the discovery of magnetic Weyl semimetals [20-23]. The interplay of magnetism and band topology may lead to the emergence of novel physical behavior, which would give new access to applications of topological states of quantum matter.

The QAH insulator and magnetic Weyl semimetal states are closely related to the band inversion under the TRS breaking. As illustrated in Fig. 1(a), regardless of whether the initial TRI systems with inversion symmetry are topological insulators with band inversion or ordinary insulators, when the TRS is broken, the exchange field leads to an inversion of one pair of spin-splitting subbands with opposite chirality at the TRI point, whereas the other pair is separated. In two-dimensional (2D) systems, the band inversion opens an energy gap characterized by a nonzero Chern number, forming the QAH insulators [10]. In three-dimensional (3D) systems, the band inversion cannot be fully gapped even with spin-orbit coupling, forming the magnetic topological semimetals (TSMs) with two-fold degenerate Weyl points or nodal lines [24]. Under the 2D limit, the degeneracy is lifted by opening an energy gap and the magnetic TSMs are transformed into the QAH insulators [20,24].

While it is believed that the QAH insulator state in magnetically doped $Bi_2Te_3$ [10-15] and $MnBi_2Te_4$ [16-19] and the magnetic Weyl semimetal state in $Co_3Sn_2S_2$ [20-23] arise from the mechanism illustrated in Fig. 1(a), the topological phase transitions driven by the TRS breaking have not been observed. In this work, we use angle-resolved photoemission spectroscopy (ARPES) to visualize the topological phase transition from a TRI ordinary insulator state to a TRS-broken TSM state in $EuB_6$, which is accompanied by the development of a long-range ferromagnetic (FM) order. Remarkably, the band crossings are the only feature at the Fermi level ($E_F$) without the existence of other Fermi surfaces (FSs), which is favorable to the emergence of exotic physics associated with the topological nodes.

## II. METHODS

Single crystals of EuB$_6$ were grown by the flux method. The ingredients with a ratio of Eu:B:Al = 1:6:570 were put into the alumina crucible with a cap and sealed into a quartz ampoule. Then the ampoule was heated to 1350 °C, kept for 10 hours, and cooled to 700 °C at a rate of 5 °C per hour. EuB$_6$ single crystals were obtained after the Al flux was removed by dissolving in sodium hydroxide solution. All the ARPES data shown were collected at the "dreamline" beamline and the 03U beamline at the Shanghai Synchrotron Radiation Facility, the SIS-HRPES beamline at the Swiss Light Source, and the BL25SU beamline at the Spring 8. We conducted complementary experiments at the 13U beamline at the National Synchrotron Radiation Laboratory at Hefei, the "Bloch" beamline, MAXIV, Sweden, the 7.0.2.1 beamline, ALS, USA, and the I05 beamline, Diamond, UK. The energy and angular resolutions were set to 15 – 70 meV and 0.1°. The samples were cleaved *in situ* and measured in ultrahigh vacuum better than 7 × 10$^{-11}$ torr.

## III. RESULTS

EuB$_6$ has the CsCl-type crystal structure with space group *Pm3m* (No. 221) [25]. As seen in Fig. 1(b), Eu atoms are located at the corners of cubic lattice and six B atoms form an octahedral cage inside the lattice. The magnetic susceptibility ($\chi$) curves in Fig. 1(f) show that EuB$_6$ is in the paramagnetic (PM) state at high temperatures. The fitting to the Curie-Weiss formula gives an effective magnetic moment $\mu_{eff}$ = 7.798 $\mu_B$/Eu$^{2+}$, which is close to the theoretical value 7.94 $\mu_B$ of local 4$f^7$ orbitals. The magnetic susceptibility starts to rise sharply at ~15 K and reaches a plateau at ~11 K in Fig. 1(f), indicating a PM to FM transition consistent with previous reports [26-28]. Specific heat measurements revealed two consecutive phase transitions at $T_M$ = 15.5 K and $T_C$ = 12.6 K [29], which were subsequently confirmed and extensively studied by various experimental techniques [30-33]. These studies indicate that isolated magnetic polarons start to form at ~30 K, and they grow in size and/or increase in number with decreasing temperature. The magnetic polarons overlap and percolate in the bulk at $T_M$, forming a ferromagnetically ordered phase coexisting with the PM phase. With further decreasing temperature, the volume fraction of the FM phase increases at the expense of the PM phase, and eventually a complete FM state is achieved at $T_C$. The adiabatic magnetization curve at 2 K in Fig. 1(g) shows that EuB$_6$ is a soft ferromagnet with an extremely low coercive field ~35 Oe.

Theoretical calculations indicate that EuB$_6$ is a semiconductor with trivial topology in the PM state [24]. Both valence band top and conduction band bottom are located at three TRI points X, Y, and Z, which are equivalent in the PM state, forming a direct band gap of ~20 meV at $E_F$ in Fig. 1(d). The valence and conduction bands near $E_F$ are mainly from the B 2$p$ and Eu 5$d$ states, respectively. When the Eu$^{2+}$ 4$f$ magnetic moments form a FM order, the inter-atomic antiferromagnetic exchange coupling between the B 2$p$ and occupied (unoccupied) Eu 4$f$ states pushes the spin-up (spin-down) valence subband upwards (downwards), while the intra-atomic FM exchange coupling between the Eu 5$d$ and occupied (unoccupied) 4$f$ states pushes the spin-up (spin-down) conduction subbands downwards (upwards) [24,34,35]. Because of the special exchange coupling, the band inversion occurs at three TRI points in the spin-up channel, while the band gap increases in the spin-down channel, as seen in Fig. 1(e). According to theoretical analysis [24], the band inversion at these TRI points cannot be fully gapped, forming Weyl points or nodal lines with two-fold degeneracy, depending on the magnetic moment direction. When the magnetic moments are aligned with the [001] direction, the mirror symmetry $\widehat{M}_z$ protects three nodal lines encircling three TRI points, respectively. When the magnetic moments are aligned with the [110] direction, the mirror symmetry $\widehat{M}_{110}$ protects one nodal line encircling the Z point while Weyl points can be found near the X and Y points. When the magnetic moments deviate from the [001] and [110] directions, no nodal lines survive because the mirror symmetries are broken, but Weyl points always exist near three TRI points and their positions move with varying the magnetic moment direction.

We first measure electronic structures in the PM state ($T$ = 20 K) on (001) cleavage surfaces of EuB$_6$ with ARPES. We observe two sets of distinct electronic structures near $E_F$ and the corresponding core level spectra have different peak structures. As seen in Fig. 2(a), the B 1$s$ (Eu 4$f$) states have single-peak (double-peak) and double-peak (single-peak) structures, respectively. As chemical environments of the atoms in the outmost layer are different from those of the atoms underneath, the atoms in the outmost layer can produce additional peaks in the core level states. We attribute the double-peak structures of the B 1$s$ and Eu 4$f$ states to two kinds of terminated surfaces with either B or Eu atoms in the outmost layer, as illustrated in Figs. 2(b) and 2(c). Recent scanning tunnelling microscopy measurements on (001) cleavage surfaces of EuB$_6$ found two kinds of distinctive topographies, which were assigned to Eu- and B-terminated surfaces, respectively [36,37]. By scanning the Eu 4$f$ and B 1$s$ core levels at different positions on the cleavage surfaces with incident light, whose spot size varies between 30 × 50 μm$^2$ and 50 × 150 μm$^2$ at various synchrotron beamlines, we can find single termination areas with only one set of electronic structures. The results were repeated in independent measurements on eight samples.

The electronic structures measured on Eu-terminated surfaces are plotted in Figs. 2(d)–2(f). We observe two electron bands near $E_F$ at $\bar{\Gamma}$ and $\bar{X}$, respectively, where $a$ is the lattice constant. Figure 2(f) shows that the bottoms of both bands have no obvious dispersion with varying photon energy ($hv$) except for the downward shift during measurements (see more details on time-dependent changes of the bands in Fig. 5 in the Appendix). The nondispersive characteristic along the surface-perpendicular momentum direction $k_z$ indicates that the bands observed on Eu-terminated surfaces are surface states. The electron band at the $\bar{X}$ point as well as its energy shift with time was also found in previous ARPES experiments [38,39], where it was speculated that the freshly cleaved surfaces could be metastable and relax with time, resulting in time-dependent band bending.

On the contrary, the $hv$-dependent data collected on B-terminated surfaces in Figs. 2(g) and 2(h) exhibit an obvious $k_z$ dispersion with a period of $2\pi/a$, which is an indication of bulk states. We observe near-$E_F$ electronic states around the X point, which show a hole band with top very close to $E_F$ in Fig. 2(i). Moreover, a shallow electron band just above the hole band is observed in some samples (see Fig. 7 in the Appendix II). The observation is well consistent with the calculated bulk electronic structure of the PM state in Fig. 1(d), and it was not captured in previous ARPES experiments [38,39]. It is expected that the near-$E_F$ electronic states can be observed at $\bar{\Gamma}$ and $\bar{X}/\bar{Y}$, which are the (001)-surface projections of three TRI points. However, we find that the hole band is only clear at ($\pi/a$, $2\pi/a$) on the second BZ boundary, while it is unclear or almost invisible at other locations, such as (0, 0), ($\pi/a$, 0), and ($2\pi/a$, 0).

When the samples are cooled down from the PM to FM state, the surface states on Eu-terminated surfaces do not change except for the downward shift with time (Fig. 6 in the Appendix), whereas the near-$E_F$ electronic states on B-terminated surfaces exhibit dramatic changes. In Fig. 3(a) one can see clear band splitting with decreasing temperature through $T_C$ on B-terminated surfaces. After the temperature-dependent measurements, we heat up the sample to 17 K and confirm that the band dispersions can be restored to the initial state in the PM state, thus excluding the possibility of time-dependent changes. The band splitting is an indication of symmetry breaking, such as the breaking of TRS or inversion symmetry. The former induces the Zeeman splitting with energy shifts while the latter induces the Rashba splitting with momentum shifts. In $EuB_6$, it is clear that the TRS is broken at low temperatures due to the long-range FM order, whereas X-ray and neutron diffraction experiments did not find any signature of a structural phase transition around $T_C$ [30,40], indicating that the inversion symmetry is preserved in the FM state. Therefore, we attribute the dramatic changes to the Zeeman splitting, which leads to energy shifts of the bands.

As seen in Fig. 3(b), there is a hole band (α) below $E_F$ at the X point at 20 K, and the band splits into two subbands ($α_1$ and $α_2$) at 5 K. The two subbands move up and down by ~200 meV, respectively, from 20 to 5 K. On the other hand, we observe that an electron band ($β_1$) move down with decreasing temperature. Due to the TRS breaking, the conduction band just above $E_F$ should also split into two subbands. One of them moves down across $E_F$, whereas the other should move up and therefore cannot be detected in the ARPES experiments. The dramatic changes in the electronic structure through $T_C$ are consistent with optical conductivity experiments, which revealed a substantial shift towards higher frequencies of the plasma edge below $T_C$, suggesting an increase of itinerant charge carriers and/or a reduction of their effective mass [29].

Notably, the band splitting results in the band inversion of two subbands $α_1$ and $β_1$ at the X point. The band inversion agrees well with the calculations in the FM state in Fig. 1(e) [24]. Moreover, the calculations indicate that two subbands $α_2$ and $β_1$ are also inverted in Fig. 1(e), while the experimental data show that they are almost degenerate at ~0.2 eV below $E_F$ in Fig. 3(b). In the experimental data both $α_2$ and $β_1$ move down significantly, whereas in the calculations only $β_1$ moves down significantly while $α_2$ moves down slightly, suggesting that the exchange coupling between the B 2$p$ and unoccupied Eu 4$f$ bands is underestimated in the calculations [24]. Despite the discrepancy, both experiment and calculation consistently confirm the band inversion of $α_1$ and $β_1$, which is critical for the magnetic TSM state in $EuB_6$.

Figures 3(c) and 3(d) show temperature-dependent momentum distribution curves (MDCs) at $E_F$ from the PM to FM state. The MDCs exhibit a single peak at 20 K, corresponding to the hole band top in the PM state. The peak starts to split at ~11 K, which can be regarded as an indication of the band inversion. We notice that the band dispersions are not well-defined between 11 K and 8 K in Fig. 3(a). The phenomenon is probably related to the phase separation between $T_M$ and $T_C$, where the FM phase formed by magnetic polarons coexists with the PM phase [29-33]. The band dispersions become clear below 8 K, probably because the complete FM state is achieved. The temperature range between 11 K and 8 K is lower than the two transition temperatures $T_M$ = 15.5 K and $T_C$ = 12.6 K determined by specific heat and magnetization measurements [29,30]. In our ARPES experiments, the actual sample temperature may be higher than that detected by the sensor because the samples are exposed to thermal radiation from the room-temperature chamber and analyzer.

We systematically investigate in-plane and out-of-plane band dispersions near the X point in the FM state. The results in Figs. 4(a)–4(f) exhibit that the band crossings at $E_F$ form a nodal-ring FS encircling the X point in the $k_x$-$k_y$ plane. As seen in Fig. 4(d), the FS is elliptical with its long axis along Γ–X. Similarly, the $hv$-dependent data in Figs. 4(g)–4(i) show a nodal-ring FS formed by the band crossings in the $k_y$-$k_z$ plane. The FS should be nearly isotropic with four-fold symmetry in the M–X–R plane, while the experimental FS in Fig. 4(g) appears to be stretched along $k_z$, which is caused by momentum broadening in the direction perpendicular to the sample surface. The FS formed by the band crossings can be regarded as the degeneracy of hole and electron FSs. de Haas–van Alphen and Shubnikov–de Haas (SdH) experiments found both hole and electron FSs with different sizes in the FM state [41,42], which could arise from the FS splitting when the chemical potential slightly deviates from the band crossings due to small non-stoichiometry. The extremal areas of the ellipsoidal FS in the ARPES data are estimated to be 278 T and 410 T in the X–M–R and Γ–X–M planes, respectively, which are ~15% smaller than those determined by the SdH experiments at 0.4 K [42]. As seen in Fig. 3(d), the Fermi wave vector continues to increase below 11 K and does not reach saturation at 5 K. The discrepancy in FS size can be attributed to the temperature difference between ARPES and SdH experiments.

Finally, we would like to discuss the effects of magnetic domains. As magnetic fields are not allowed in the ARPES experiments, magnetic domains must exist in the soft ferromagnet EuB$_6$. Because of the small magnetic anisotropy in EuB$_6$, the typical domain size at zero field was estimated to be on the order of 1 μm [30], which is much smaller than the spot size of incident light, suggesting that the ARPES results should be a mixture of electronic structures of multiple domains with various magnetic moment directions. As the magnetic moment direction determines the band crossings at $E_F$, specific topological states cannot be distinguished in the mixed electronic structure of multiple domains. On the other hand, the effects of magnetic moment direction on the overall band dispersions are negligible [24]. Therefore, despite the existence of magnetic domains, clear band dispersions can be observed when the samples enter the complete FM state at low temperatures.

## IV. OUTLOOK

Our ARPES data have demonstrated that the TRS breaking results in a topological phase transition with the band inversion at the X point in EuB$_6$, which is well consistent with the band calculations [24]. According to theoretical analysis, the band inversion at the TRI point cannot be fully gapped, forming Weyl points or nodal lines depending on the magnetic moment direction. As a soft ferromagnet, it is easy to align the magnetic moments with external magnetic fields, which would be favorable to tune the topological states and relevant physical properties. Compared with the other experimentally discovered magnetic TSMs [21,23,43], the distinguishing characteristic in EuB$_6$ is that the band crossings are the only feature at $E_F$ without the existence of other FSs, and therefore their energy positions are restricted at $E_F$ by electron-hole compensation in stoichiometric samples.

Band calculations indicate that the band inversion occurs in the spin-up channel whereas the band gap remains in the spin-down channel, suggesting that EuB$_6$ is also a half metal with a full spin polarization at $E_F$ in the FM state [24,33,44]. Our experimental data are consistent with the predicted half-metallic electronic structure. However, the expected full spin polarization appears to be in conflict with Andreev reflection spectroscopy results showing a spin polarization of ~50 %, which is attributed to the scenario that one band is fully spin-polarized while the other is unpolarized [45]. It has been argued that the half-metallic state can be tuned into an incomplete spin-polarized state by slight doping [34]. Further studies are required to resolve the disagreement.

Since Weyl points can only be defined in 3D momentum space, they will be removed by opening an energy gap under the 2D limit. If the band order is maintained during the crossover from 3D to 2D, the Weyl semimetal state can be transformed into the QAH insulator state. Band calculations have predicted that the QAH insulator state can be achieved under the 2D limit in EuB$_6$, where the band crossings open an energy gap [24]. As there is no interference from other bands, the QAH insulator state has a global gap at $E_F$ in the density of states. It is expected that the gap size reaches a maximum of ~100 meV at a thickness of 30 – 40 Å in thin films of EuB$_6$ [24]. These outstanding properties make EuB$_6$ a promising system to study the physical behavior associated with a range of magnetic topological states in both 3D crystals and 2D thin films.

## Acknowledgements

T.Q. thanks Xi Dai for the initial idea. We thank Zhicheng Jiang, Zhenyu Yuan, Takayuki Muro, and Seigo Souma for technical assistance. This work was supported by the Ministry of Science and Technology of China (2016YFA0300600, 2019YFA0308602, 2016YFA0401000, and 2017YFA0302901), the National Natural Science Foundation of China (U1832202, 11874422, 11888101, U2032204, 12004416, and 11974395), the Chinese Academy of Sciences (QYZDB-SSW-SLH043, XDB33020100, and XDB28000000), the Fundamental Research Funds for the Central Universities, and the Research Funds of Renmin University of China (19XNLG18 and 18XNLG14), the Beijing Municipal Science and Technology Commission (Z171100002017018 and Z181100004218005), the Beijing Natural Science Foundation (Z180008), the K. C. Wong Education Foundation (GJTD-2018-01), the Center for Materials Genome, and the Users with Excellence Program of Hefei Science Center CAS (2019HSC-UE001). N.P. thanks the support from the Villum Fonden, for the NEED project (00027993). M.S. was supported by the Sino-Swiss Science and Technology Cooperation (IZLCZ2-170075) and the Swiss National Science Foundation (200021_188413).

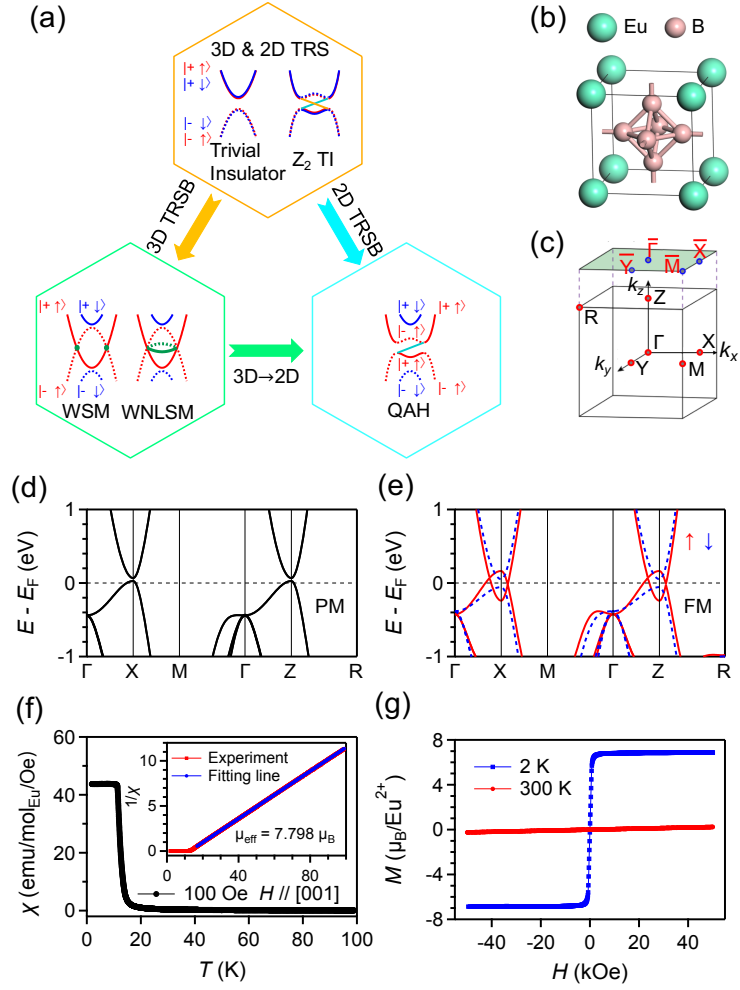

Fig. 1. Calculated electronic structures and magnetic properties of EuB$_6$. (a) Schematic plots of emergent topological phases with TRS breaking (TRSB) in centro-symmetric 3D and 2D systems. TI, WSM, and WNLSM are the abbreviations of topological insulator, Weyl semimetal, and Weyl nodal-line semimetal. "+" and "−" represent opposite parity of the subbands. "↑" and "↓" represent spin up and spin down. (b) Crystal structure of EuB$_6$. (c) Bulk BZ and (001) surface BZ. (d),(e) Calculated electronic structures along high-symmetry lines in the PM and FM states, respectively. Magnetic moments are defined to be along the [001] direction in the FM state. The calculated results are adopted from Ref. [24]. (f) Temperature dependence of magnetic susceptibility. Inset: Temperature dependence of inverse susceptibility. The blue line is a fitting to the Curie-Weiss formula. (g) Magnetic field dependence of magnetization at 2 K (blue) and 300 K (red) with magnetic fields parallel to the [001] direction.

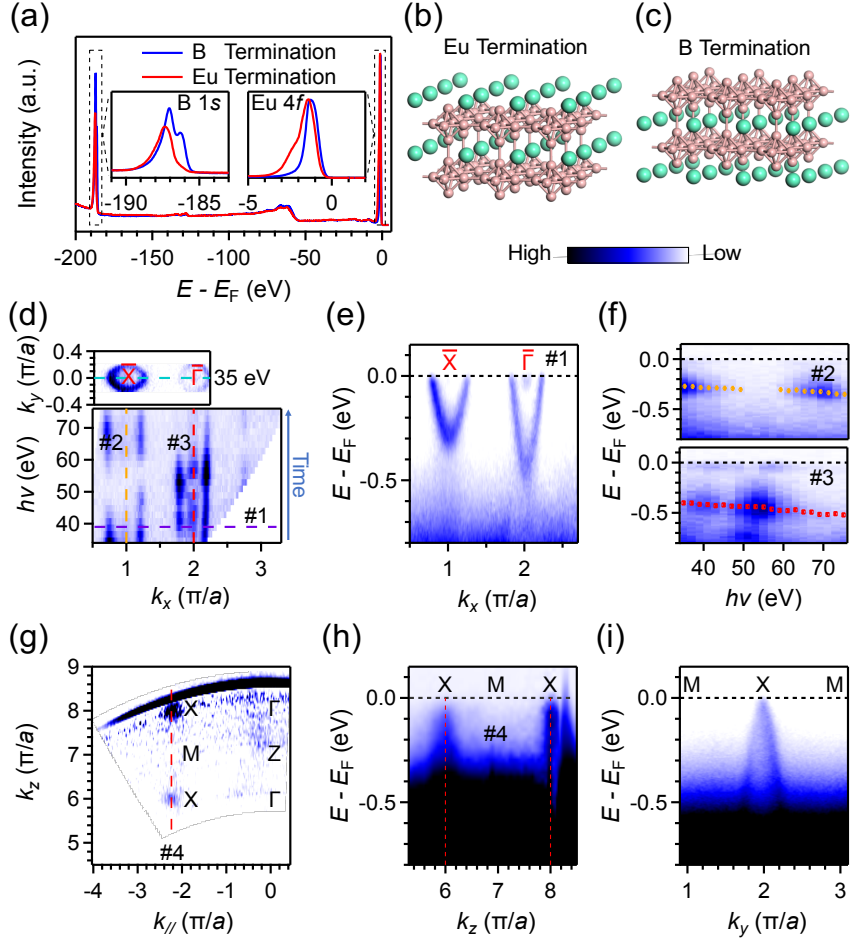

Fig. 2. Electronic structures measured on Eu- and B-terminated surfaces in the PM state. (a) Core-level photoemission spectra showing characteristic peaks of the B 1$s$ and Eu 4$f$ orbitals in the insets. (b),(c) Schematic plots of (001) cleavage surfaces with Eu and B terminations, respectively. (d) Top: Intensity plot of the ARPES data at $E_F$ with $h\nu$ = 35 eV. Bottom: Intensity plot of the ARPES data at $E_F$ collected in a range of photon energies from 34 to 76 eV. (e) Intensity plot of the ARPES data along cut #1. (f) Intensity plots of the ARPES data along cuts #2 and #3, showing photon energy and time dependence of the band bottoms. (g) Intensity plot of the ARPES data at $E_F$ collected along the direction from (0, 0) to (-$\pi/a$, 2$\pi/a$) in a range of photon energies from 59 to 151 eV. The inner potential $V_0$ = 17.73 eV and the lattice constant $a$ = 4.187 Å. (h) Intensity plot of the ARPES data along cut #4. (i) Intensity plot of the ARPES data along M–X in the $k_x$-$k_y$ plane with $k_z = 8\pi/a$. The data in (d)-(f) and (g)-(i) were collected on Eu- and B-terminated surfaces, respectively. All the data in Fig. 2 were collected at 20 K.

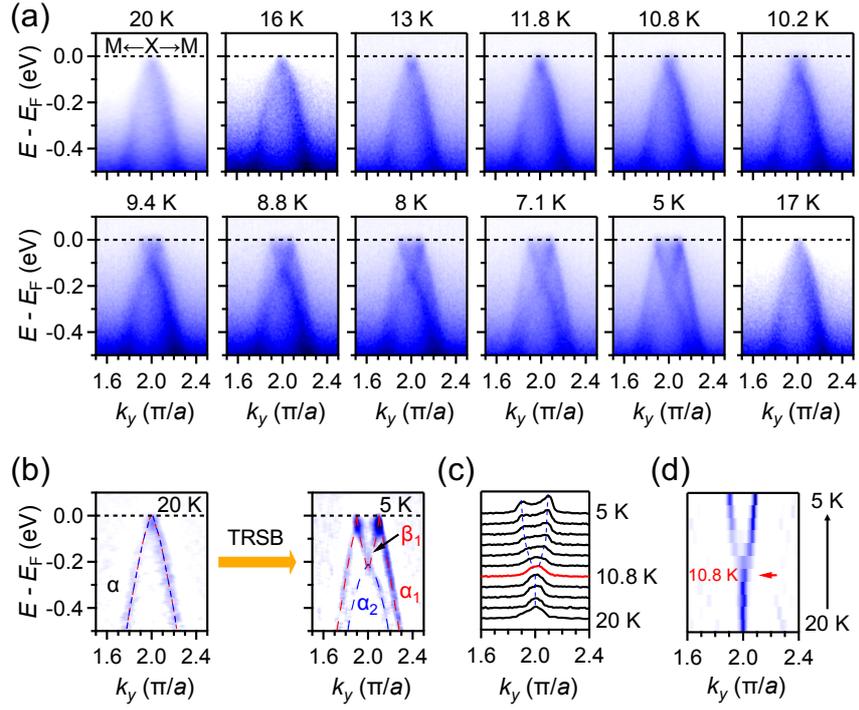

Fig. 3. Temperature dependence of band dispersions on B-terminated surfaces. (a) Intensity plots of the ARPES data along M–X at different temperatures between 20 K and 5 K. (b) 2D curvature intensity plots of the data in (a) at 20 K and 5 K. Dashed lines are guide to eyes of the band dispersions. (c) MDCs at $E_F$ of the ARPES data in (a). (d) Curvature intensity plot of the data in (c). All the data in Fig. 3 were collected with $h\nu$ = 135 eV.

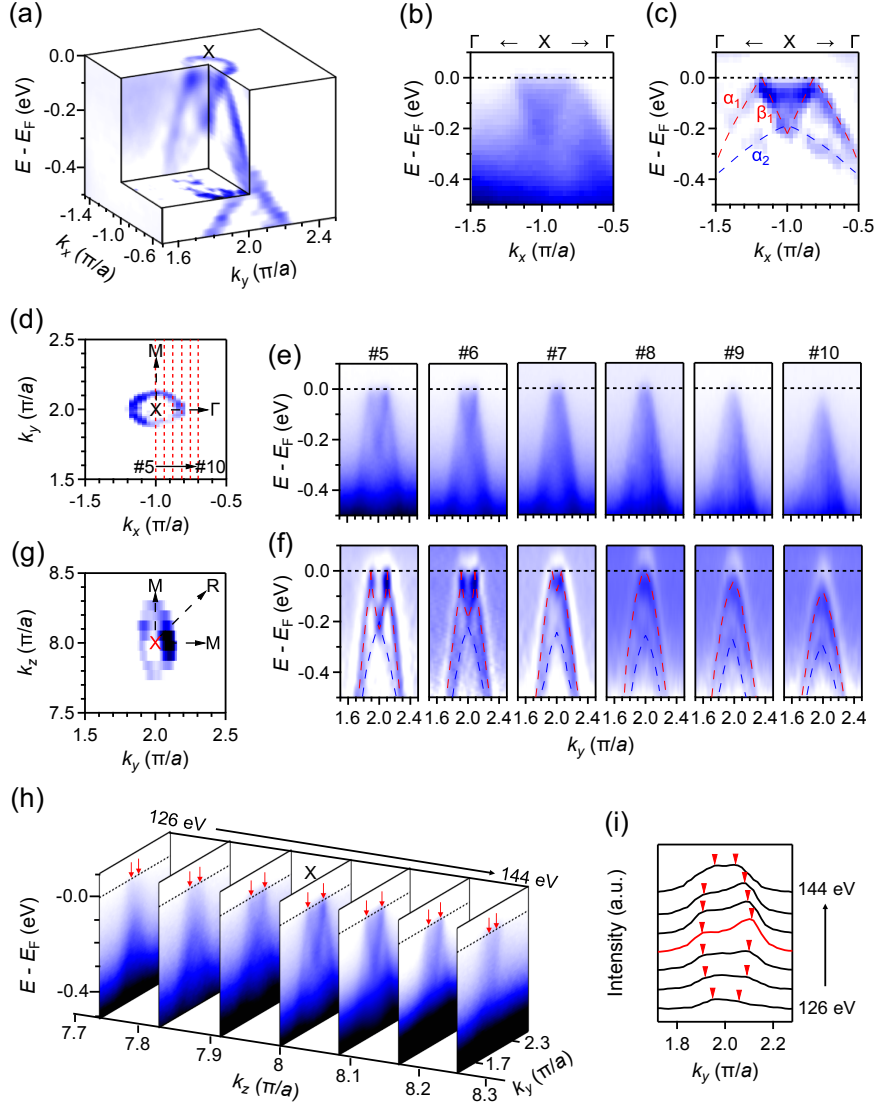

Fig. 4. In-plane and out-of-plane band dispersions on B-terminated surfaces in the FM state. (a) 3D curvature intensity plot of the ARPES data around the X point. (b) Intensity plot of the ARPES data along Γ–X. (c) 2D curvature intensity plot of the data in (b). Dashed lines are guide to eyes of the band dispersions. (d) Curvature intensity plot of the ARPES data at $E_F$ around the X point in the $k_x$-$k_y$ plane. (e) Intensity plots of the ARPES data along cuts #5 to #10. (f) Curvature intensity plots of the ARPES data in (e). (g) Curvature intensity plot of the ARPES data at $E_F$ around the X point in the $k_y$-$k_z$ plane. (h) Intensity plots of the ARPES data along M–X collected in a range of photon energies from 126 to 144 eV. (i) MDCs at $E_F$ of the ARPES data in (h). The data in (a)–(f) were collected with $h\nu$ = 135 eV. All the data in Fig. 4 were collected at 4.5 K.

# APPENDIX I: TIME DEPENDENCE OF BAND DISPERSIONS MEASURED ON Eu-TERMINATED SURFACES

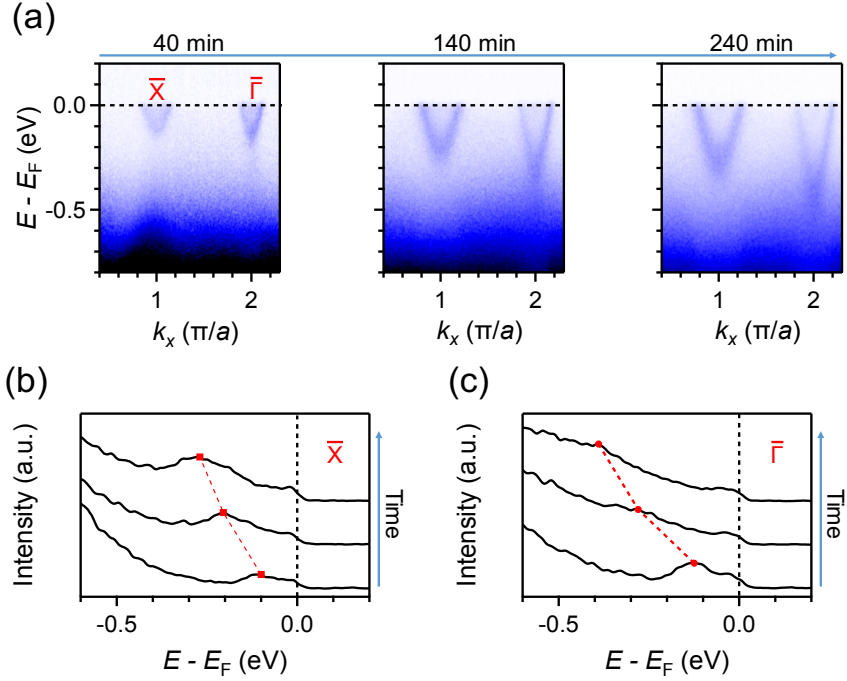

Fig. 5. Time dependence of band dispersions measured on Eu-terminated surfaces at 60 K. (a) Intensity plots of the ARPES data collected at 40, 140, and 240 minutes after the cleavage with $h\nu = 38$ eV. (b),(c) Energy distribution curves at the $\bar{X}$ and $\bar{\Gamma}$ points, respectively, showing energy shifts with time of the band bottoms at 60 K.

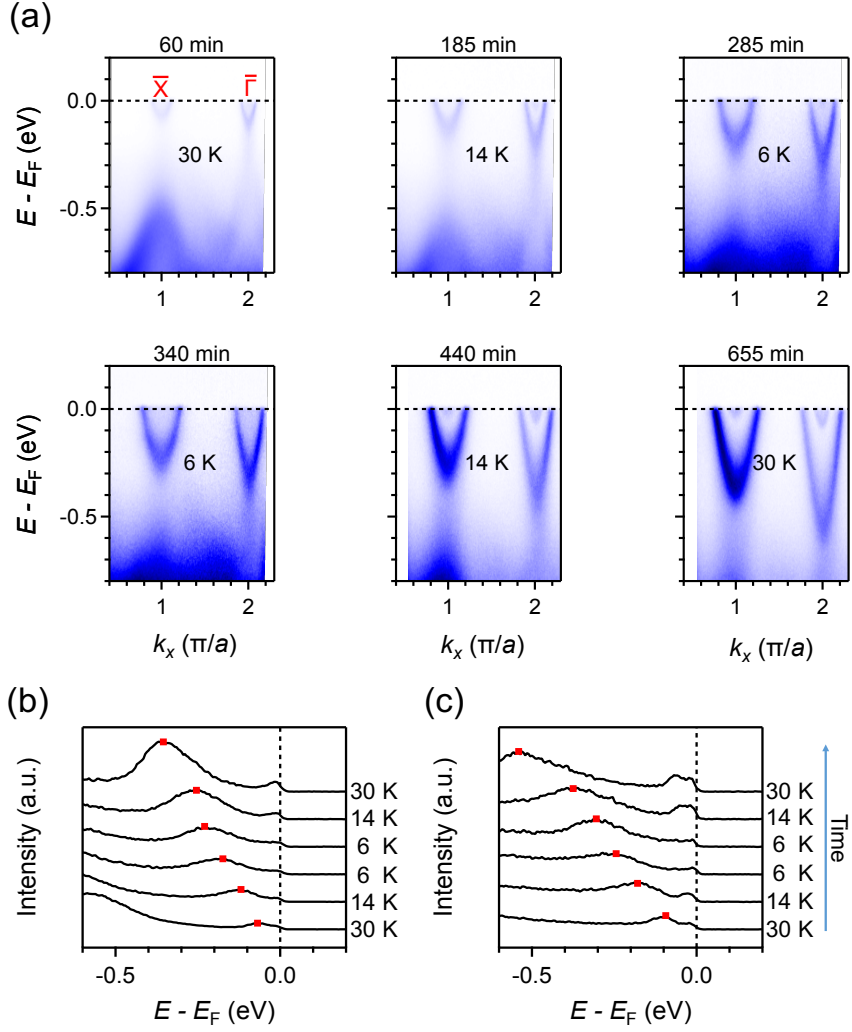

Fig. 6. Time dependence of band dispersions measured on Eu-terminated surfaces under temperature cycling. (a) Intensity plots of the ARPES data collected under temperature cycling in time sequence. (b),(c) Energy distribution curves at the $\bar{X}$ and $\bar{\Gamma}$ points, respectively, showing that energy shifts of the band bottoms are related to time rather than temperature.

# APPENDIX II: SURFACE DOPING EFFECTS ON Eu- AND B-TERMINATED SURFACES

Since the cleavage produces polar (001) surfaces with either Eu or B termination, it is expected that the electronic states on the polar surfaces suffer from surface doping. As B has a stronger electronegativity than Eu, Eu has a positive valence state in $EuB_6$. The surface doping would cause the electronic states on Eu-terminated surfaces to be more *n*-type compared with those in the bulk, which could explain the electron pockets observed on Eu-terminated surfaces in Figs. 2(d) and 2(e).

As B has a negative valence state in $EuB_6$, the surface doping would cause the electronic states on B-terminated surfaces to be more *p*-type compared with those in the bulk. However, the electronic states observed on B-terminated surfaces appear to be either nearly neutral in Fig. 2(i) or slightly *n*-type in Fig. 7, which seems to contradict the expectation. One possible scenario is that the electronic states on B-terminated surfaces are coherent with the bulk states since they exhibit an obvious $k_z$ dispersion, which significantly reduces the surface doping effects. Another possibility is that the bulk states are slightly *n*-type because Hall effect measurements show a negative Hall coefficient [46-48], and the surface hole doping on B-terminated surfaces tends to neutralize the intrinsic *n*-type carriers.

The electronic states on B-terminated surfaces are observed to be nearly neutral in six samples and slightly *n*-type in two samples, indicating small fluctuations in the stoichiometry of the samples. In addition, the data in Fig. 7 confirm the semiconducting electronic structure with a narrow band gap in the PM state, which is consistent with the calculations in Fig. 1(d).

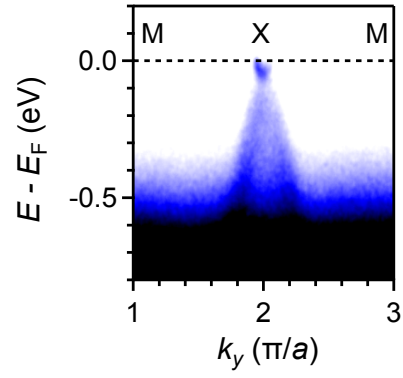

Fig. 7. Intensity plot of the ARPES data along M–X in the $k_x$-$k_y$ plane with $k_z = 8\pi/a$ at 22 K, showing slightly *n*-type electronic states on B-terminated surfaces of some samples.